\documentclass{ptapap}
\author{Andrzej Pigulski}[IAUWr]
\author{Henryk Cugier}[IAUWr]
\author{Gerald Handler}[CAMK]
\author{Refilwe Kgoadi}[UoJ,CoS]
\author{the BRITE Team}
\affil[IAUWr]{Instytut Astronomiczny, Uniwersytet Wroc{\l}awski, Wroc{\l}aw, Poland}
\affil[CAMK]{Nicolaus Copernicus Astronomical Center, Warszawa, Poland}
\affil[UoJ]{Department of Physics, University of Johannesburg, South Africa}
\affil[CoS]{College of Science and Engineering, James Cook University, Douglas, Australia}
\title{BRITE view of $\sigma$~Scorpii, $\beta$~Cephei-type star studied for over a century}

\begin{document}
\maketitle
\begin{abstract}
Preliminary results of the analysis of the combined space-based BRITE and SMEI, and ground-based Str{\"o}mgren photometry are presented. The BRITE data allowed to find seven $p$ and three $g$ modes in the frequency spectrum of this star; only four $p$ modes were known in this star prior to this study. The first results of seismic modelling are also presented.
\end{abstract}

The bright star $\sigma$~Scorpii (HD~147165, $V=$ 2.89\,mag, B1\,III) is a hierarchical quadruple system consisting of B-type stars. 
The closest Aa-Ab pair is a double-lined (SB2) spectroscopic binary \citep{1952AJ.....57Q..16L,1955ApJ...122..122S} in a 33-day orbit.
The variability of radial velocities of $\sigma$~Sco was first detected more than a century ago by \cite{1904ApJ....20..146S}. Later spectroscopic observations 
revealed not only the SB2 nature of the Aa-Ab pair, but also multiperiodic pulsations of its primary (Aa) component. 
Following the discovery of the photometric variability of $\sigma$~Sco by \cite{1951MNRAS.111..339H}, the star was also studied photometrically. 
The analysis of the ground-based data revealed the presence of four $p$ modes 
\citep{1984MNRAS.211..297J}. The three strongest modes were also detected in the variability of radial velocities and line profiles \citep{2014MNRAS.442..616T}. 

\begin{figure}
\centering
\includegraphics[width=0.7\textwidth]{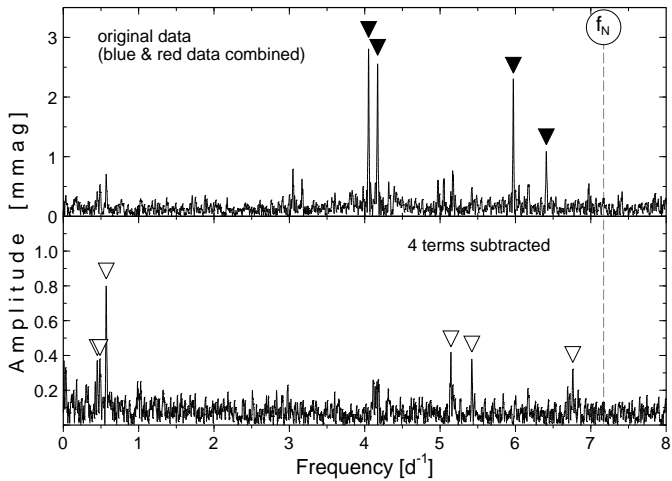}
\caption{Frequency spectrum of the combined blue and red BRITE data of $\sigma$~Sco. Top: Frequency spectrum of the original data. The peaks corresponding to the frequencies of the four known modes are marked with inverted triangles. Bottom: the same after subtracting the four strongest terms. The newly discovered modes are marked by open inverted triangles. The vertical line labeled with $f_{\rm N}$ marks the orbital sampling Nyquist frequency.}
\label{fig:ssco}
\end{figure}

The star was observed by BRITE-Constellation \citep{2014PASP..126..573W,2016PASP..128l5001P} as one of 26 BRITE targets in the Scorpius I field. The observations were made between March 19th and August 29th, 2015, with four out of five working BRITE satellites: blue-filter BRITE-Austria (BAb) and BRITE Lem (BLb) and red-filter UniBRITE (UBr) and BRITE Heweliusz (BHr). The observations resulted in 125\,681 data points obtained by means of the pipeline presented by \cite{2017A&A...605A..26P}. The data corrected for instrumental effects were subject of time-series analysis, which resulted in the detection of 11 periodic terms in the combined blue and red data (Fig.~\ref{fig:ssco}). The final solution includes four known modes, six new ones, and the harmonic of the main mode. Three new modes have frequencies in the $p$-mode range, the remaining three are likely $g$ modes. All new modes have amplitudes below 1\,mmag.
The discovery of $g$ modes in $\sigma$~Sco makes the star another bright hybrid $\beta$~Cep/SPB object.
The hybridity of this kind has been revealed by BRITE-Constellation for many other `classical' bright $\beta$~Cep-type stars, for example $\beta$~Cen \citep{2016A&A...588A..55P}, $\theta$~Oph and $\kappa$~Sco (Walczak et al., in prep.). The discovery of new modes allows also for a more detailed seismic modeling.

The star was also observed from space by means of SMEI experiment onboard Coriolis satellite \citep{2004SoPh..225..177J}. The data cover about 8 years, between 2003 and 2010. The frequency spectrum of the SMEI data allows for a detection of only the four strongest terms.

One year before BRITE photometry was obtained, $\sigma$~Sco was observed by means of the Automatic Photometric Telescope (APT) in Fairborn Observatory, Arizona, and with 50-cm SAAO telescope using Str{\"o}mgren $u$, $v$, and $y$ passbands. The resulting frequency spectrum shows only the three strongest terms. As already indicated by \cite{2014IAUS..301..417H}, the observations revealed a significant drop in amplitude of all pulsation modes known at the time. The space-based data confirm a roughly fourfold decrease of the amplitude of the main mode between 1984 and 2003.

With 10 modes detected in the BRITE data of $\sigma$~Sco, we attempted seismic modeling. It was based on the evolutionary models that take into account stellar rotation according to \cite{1998A&A...334..911S}; see \cite{2016A&A...588A..55P} for more details. Assuming that the strongest mode with frequency $f_1=4.051$\,d$^{-1}$ is radial \citep{1992AcA....42..191C}, we obtained the following preliminary results:
\begin{itemize}
\item For the 14\,M$_\odot$ model, the two $g$ modes are unstable only if $f_{\rm rot} > 0.266$\,d$^{-1}$ ($V_{\rm rot} > 120$\,km/s). This implies a small inclination, consistent with the rotation axis perpendicular to the orbital plane.
\item The central hydrogen content ranges between 0.172 and 0.243, that is, the star is advanced in its main-sequence evolution.
\item The two modes with the highest frequencies cannot be explained with the 14\,M$_\odot$ models and $l = 0-2$. The modes have either $l > 2$ or originate in the secondary.
\item Some models fit three consecutive radial modes. However, the one with the highest frequency (6.761\,d$^{-1}$) is always stable in the models.
\item In some models, two unstable modes are identified as  $(l,m) =$ (1,0), and (2,0). One of them has observed amplitude comparable to the radial mode.
\end{itemize}

A detailed analysis of the presented data and seismic modelling will be published elsewhere.
\acknowledgements{Based on data collected by the BRITE Constellation satellite mission, designed, built, launched, operated and supported by the Austrian Research Promotion Agency (FFG), the University of Vienna, the Technical University of Graz, the Canadian Space Agency (CSA), the University of Toronto Institute for Aerospace Studies (UTIAS), the Foundation for Polish Science \& Technology (FNiTP MNiSW), and National Science Centre (NCN). The operation of the Polish BRITE satellites is secured by a SPUB grant of the Polish Ministry of Science and Higher Education (MNiSW). The work was supported by the BRITE PMN grant 2011/01/M/ST9/05914. AP and HC acknowledge support from the National Science Centre (NCN) grant No.~2016/21/B/ST9/01126. GH acknowledges support from the NCN grant No.~2015/18/A/ST9/00578.}

\end{document}